\documentclass[pre,pdftex,twocolumn,byrevtex,superscriptaddress]{revtex4}
\usepackage{amsmath,epsfig,enumerate,graphicx}
\usepackage{verbatim}
\usepackage{amstext}
\usepackage{ifthen}
\newboolean{twocolswitch}
\newboolean{revtexswitch}

\newcommand{\req}[1]{(\ref{#1})}

\setboolean{twocolswitch}{true}
\setboolean{revtexswitch}{true}

\begin{document}

\title{
A Touch of Sleep:\\ 
Biophysical Model of Contact-mediated Dormancy of Archaea by Viruses
}

\author{
  \firstname{Hayriye}
  \surname{Gulbudak}
  }
\affiliation{
   	School of Biology,
	Georgia Institute of Technology,
	Atlanta, GA, USA
	}
\author{
  \firstname{Joshua S.}
  \surname{Weitz}
  }
\email{jsweitz@gatech.edu}
\homepage{http://ecotheory.biology.gatech.edu}
\affiliation{
   	School of Biology,
	Georgia Institute of Technology,
	Atlanta, GA, USA
	}
\affiliation{
   	School of Physics,
	Georgia Institute of Technology,
	Atlanta, GA, USA
}

\date{\today}

\begin{abstract}
The canonical view of the interactions between viruses and
their microbial hosts presumes that changes in host and virus
fate require the initiation of infection of a host by a virus.  That is,
first virus particles diffuse randomly outside of host cells,
then the virus genome enters the target host cell, and only then
do intracellular dynamics and regulation of virus
and host cell fate unfold.  Intracellular dynamics
may lead to the death of the host cell and release of viruses,
to the elimination of the virus genome through 
cellular defense mechanisms, or the integration of 
the virus genome with the host as a chromosomal or extra-chromosomal 
element.  Here we revisit this canonical view, inspired by 
recent experimental findings of Bautista and colleagues (mBio, 2015)
in which
the majority of target host cells can be induced into a dormant state
when exposed to either active
or de-activated viruses, even when viruses are present at low relative titer.  
We propose that both the qualitative
phenomena and the quantitative time-scales of dormancy induction
can be reconciled given the hypothesis that
cellular physiology can be altered by \emph{contact}
on the surface of host cells rather than strictly by \emph{infection}.
We develop a biophysical model of contact-mediated dynamics involving
virus particles and target cells.  We show how in this model
virus particles can catalyze -- extracellularly --
cellular transformations amongst many cells, even if they ultimately
infect only one (or none). We discuss 
implications of the present biophysical model 
relevant to the study of virus-microbe interactions more generally.
\end{abstract}

\maketitle

\section{Introduction}
Dormancy is ubiquitous in microbial systems. One prominent
example is that of bacterial persistence, in which
cells undergo a rapid physiological change characterized
by slowed or even halted growth as well as decreased susceptibility to 
antibiotics ~\citep{balaban_2004,lewis2007persister,lewis_annrev2010}.  Another example is that of microbial
``seed banks''~\citep{lennon2011microbial}, in which individual
microbes undergo a long period of stasis in which growth is halted.
These cells can then restart growth given changes in environmental
conditions.  A third example is the starvation-dependent
division of division of the bacteria \emph{Sinorhizobium meliloti}
into distinct daughter phenotypes, one that is better suited
to long-term starvation~\citep{ratcliff_2010}.  
Long-standing theory~\citep{cohen_1966} and recent experiments~\citep{beaumont_2009} support the consensus that
varying phenotypes - whether by stochastic bet hedging or via phenotypic
plasticity - is an evolutionarily favorable strategy in the face
of uncertainty in environmental selection pressures~\citep{kussell_sci2005}.
One of the selection pressure that microbial cells face is the possibility
of infection and lysis by viruses. Indeed, it has been suggested
that \emph{E.~coli} may enter the persistence state as a route
to diminish, temporarily, the ability of viruses to eliminate
a local population~\citep{pearl_2008}. Nonetheless,
the transition between active growth and persistence was
not proposed to be a function of virus-host interactions.

Here, we propose a biophysical model of virus-induced dormancy
of microbial host cells.  Our model is inspired by
recent empirical findings by Bautista and colleagues~\citep{bautista_inpress}
of the interactions between
the archaeon \emph{Sulfolobus islandicus} and the dsDNA 
fusellovirus \emph{Sulfulobus} spindle shaped virus (SSV9).  
\emph{S.~islandicus} is
a globally distributed archaeon, commonly found in hot spring ecosystems.
\emph{S.~islandicus} is also a model system
for studying the eco-evolutionary basis for diversity in 
archaea~\citep{zhang2013sulfolobus,ISI:000282210700006,reno_2009,ISI:000262515800015,prangishvili_2006}.  
\begin{figure*}[t!]
\begin{center}
\includegraphics[width=0.8\textwidth]{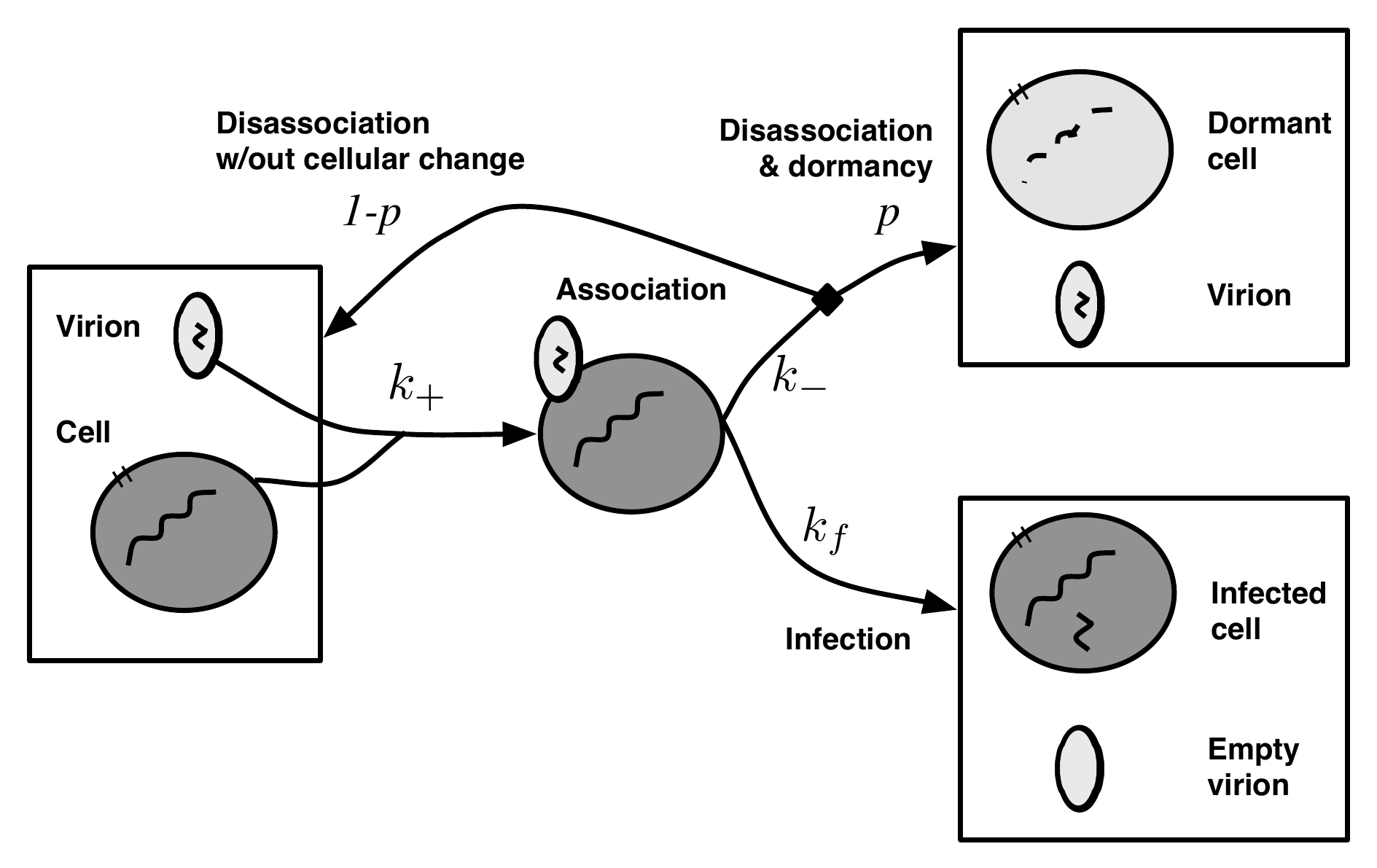}
\caption{\label{fig.model_schematic} Schematic representation of a
biophysical model of virus-host interactions
including susceptible microbial cells ($S$), complexes of
a cell with a virus particle ($C$), infected cells ($I$),
dormant cells ($D$) and free viruses ($V$).  All populations
are tracked in terms of densities of particles/ml.}
\end{center}
\end{figure*}
A recent study of the interactions
between the host strain \emph{S.islandicus} RJW002 and the virus SSV9
were recently shown to lead to a novel population-level
outcome: ``challenge of RJW002 
with SSV9 induced a population-wide stasis or dormancy response, 
where the majority of cells are viable but not actively growing''~\citep{bautista_inpress}.   
Dormant cells appear ``empty'' without coherent
intracellular structure in contrast
to normal cells.   Dormant cells can then reorganize and revert
to actively growing cells. In the experiment, viruses were introduced
at low concentrations relative to that of hosts.
Yet, after 24 hrs, nearly 100\% of cells were classified
as dormant~\citep{bautista_inpress}.  In other words:
there was a large-scale amplification
in the number of dormant cells at the end of the 24 hr period
vs.~the number of viruses at the start of the experiment.  
Further, in a follow-up experiment, nearly 100\%
of cells initiated dormancy even when the host was
exposed to \emph{de-activated} viruses at low relative concentration.

Bautista and
colleagues~\citep{bautista_inpress} highlight the potential role
of dormancy as a strategy to survive viral infection and lysis. 
These experiments
also raise the possibility that contact between viruses particles and
host surface may be sufficient to initiate a large-scale
physiological response, both at the cellular and population-scales.
The model we propose here focuses on the early dynamics of
virus-host interactions.  We assume that
viruses can ``contact'' host cells reversibly.  
If most contacts between virus particles and host cells do not lead to infection,
then there can be a broad dynamic regime in which nearly all of
the target host cells enter dormancy even when there are
far fewer viruses than hosts.  We reach this conclusion
by first proposing
a biophysical model of contact-mediated dormancy.  Next, we solve
the model and identify critical transitions between distinct
qualitative regimes, including a broad regime of dormancy-enhancement.
Finally, we apply the model to the experimental conditions
underlying the recent study of RJW002 and SSV9 and show that
a contact-model of dormancy is compatible with observations.
We discuss the implications of the current work for updating the
dominant virus-host infection paradigm to include renewed focus on the
role of virus particles in stimulating changes amongst
microbes and microbial populations.

\section{Results}

\subsection{Biophysical model of contact-mediated dormancy
of hosts by viruses}

We propose a nonlinear dynamics model of virus-host interactions 
(see Figure~\ref{fig.model_schematic}).
This model describes the early dynamics of interactions involving
viruses and hosts and the initiation of either dormancy
or an active infection. 
Consider an environment containing susceptible cells, $S$,
and free virus particles, $V$.  Free viruses can contact cells
forming a complex, $C$, given a diffusion-limited contact rate of
$k_+$ cells/(ml$\cdot$hrs).  
The use of the term ``complex'' suggests an
analogy to models of enzyme kinetics.  The complex is reversible.
The disassociation rate is $k_-$ cells/(ml$\cdot$hrs)
and the infection rate is $k_f$ hrs$^{-1}$.
If disassociation takes place, then the virus
is released back into the environment.  We assume that
disassociation may also induce a cellular transformation leading
to dormancy with probability $p$.  If infection takes place,
then the virus genome enters the host cell, leading
to an actively infected cell.  The dynamics of this model
can be written as:
\begin{eqnarray}
\label{eq.dmodel}
\begin{split}
\textrm{Susceptible~ } \frac{\mathrm{d} S}{\mathrm{d} t} &= -\overbrace{k_+ SV}^{\textrm{contact}}+\overbrace{(1-p)k_-C}^{\textrm{disassociation}} \\
\textrm{Complex~ } \frac{\mathrm{d} C}{\mathrm{d} t} &= \overbrace{k_+ SV}^{\textrm{contact}}-\overbrace{k_-C}^{\textrm{disassociation}}-\overbrace{k_fC}^{\textrm{infection}} \\
\textrm{Dormant~ } \frac{\mathrm{d} D}{\mathrm{d} t} &= \overbrace{p k_-C}^{\textrm{contact-mediated dormancy}} \\
\textrm{Infected~ } \frac{\mathrm{d} I}{\mathrm{d} t} &= \overbrace{k_fC}^{\textrm{infection}} \\
\textrm{Free viruses~ } \frac{\mathrm{d} V}{\mathrm{d} t} &= -\overbrace{k_+ SV}^{\textrm{contact}} +\overbrace{k_-C}^{\textrm{disassociation}} 
\end{split}
\end{eqnarray}

The system can be reduced in complexity. 
First there is a constraint that $S_0 = S(t)+C(t)+D(t)+I(t)$, because the dynamics
in the model track the transformation of an initial population
of $S_0$ susceptible cells into four different states: susceptible,
complex, dormant, and infected.  
There is another constraint that $V_0=C(t)+I(t)+V(t)$, because the dynamics
track the transformation of an initial population
of $V_0$ virus genomes into three different states: in free
viruses, temporarily bound with cells, and injected into hosts.
Finally, when contact occurs rapidly, then we can use a standard assumption
in enzyme kinetics theory and presume that the concentration of $C$
rapidly equilibriates (for example, see the
Appendix of ~\cite{alon_2007}).  
This is a standard approach to analyzing models characterized
by fast-slow dynamics.  Here, we assume that
the change in $C$ is relatively fast when compared to
other state variables. In the fast limit, then $C(t)=\frac{k_+SV}{k_-+k_f}$.
This approximation
is referred to as quasi-steady state approximation (QSSA).
Substituting the QSSA equilibrium for the concentration
of the complex yields the following reduced system:
\begin{eqnarray}
\label{eq.dmodel_2}
\begin{split}
\frac{\mathrm{d} S}{\mathrm{d} t} &= -SV\left(\frac{k_fk_++pk_-k_+}{k_-+k_f}\right) \\
\frac{\mathrm{d} V}{\mathrm{d} t} &= -SV\left(\frac{k_+k_f}{k_-+k_f}\right)
\end{split}
\end{eqnarray}

We can then identify the following control parameters: the
conditional probability of infection given contact, 
$q=\frac{k_f}{k_-+k_f}$, 
the effective adsorption rate, $\phi=qk_+$,
and the ratio of dormancy induction to infection, $\delta = \frac{pk_-}{k_f}$.
Using these control parameters, we can rewrite the model as:
\begin{eqnarray}
\frac{\mathrm{d} S}{\mathrm{d} t} &=& -\phi SV(1+\delta) \label{eq.dsdt} \\
\frac{\mathrm{d} V}{\mathrm{d} t} &=& -\phi SV \label{eq.dvdt}
\end{eqnarray}
This model can be interpreted as follows.  The density of susceptible
hosts decreases at a rate proportional to the densities of virus and
susceptible host populations (Eq.~\req{eq.dsdt}). The proportionality constant is $\phi$, the adsorption rate,
multiplied by an enhancement factor of $(1+\delta)$, where $\delta$
is the number of dormant cells induced for each infected cell produced. 
The enhancement factor arises due to the fact that susceptible
hosts can become infected or enter dormancy due to interactions
with viruses.  
The number of free viruses decreases at a rate proportional to the densities of virus and susceptible host populations (Eq.~\req{eq.dvdt}). 
The proportionality constant
in that case is $\phi$, the adsorption rate, because that is the means by which
free viruses are removed from the medium. 

The model does not include birth and death of hosts,
the lysis of hosts by viruses, nor the decay of virus
particles. As such, the model describes
\emph{early} dynamics of virus-host interactions. This focus
is in contrast to models of virus and host dynamics
mediated by density-dependent infection and lysis~\citep{levin_1977,weitz_2005,smith_2012,childs_2012}.

\subsection{Qualitative regimes of dormancy induction}
Eqs.~\req{eq.dsdt}--\req{eq.dvdt} can be solved analytically (see
Appendix~\ref{app.analytics}), yielding:
\begin{eqnarray}
S(t) &=& \frac{\Omega}{1-(1+\delta){\cal{M}}_0e^{-\phi \Omega t}}\label{eq.s_of_t}\\
V(t) &=& \frac{{\cal{M}}_0 \Omega e^{-\phi\Omega t}}{1-(1+\delta){\cal{M}}_0e^{-\phi \Omega t}}\label{eq.v_of_t}
\end{eqnarray}
where $\Omega=S_0-(1+\delta)V_0$ and ${\cal{M}}_0=V_0/S_0$.
These solutions hold so long as $\Omega\neq 0$.  When $\Omega=0$ then
\begin{eqnarray}
S(t) &=& \frac{S_0}{1+S_0\phi t}  \\
V(t) &=& \frac{V_0}{1+V_0\phi(1+\delta)t}   
\end{eqnarray}
The system dynamics have qualitatively different
behaviors for $\Omega>0$ and for $\Omega<0$ (see Table~\ref{tab.asymp}).  
Therefore $\Omega$ acts as a \emph{critical} parameter, both in a
biological and dynamical systems sense.

\begin{figure}[t!]
\begin{center}
\includegraphics[width=0.95\columnwidth]{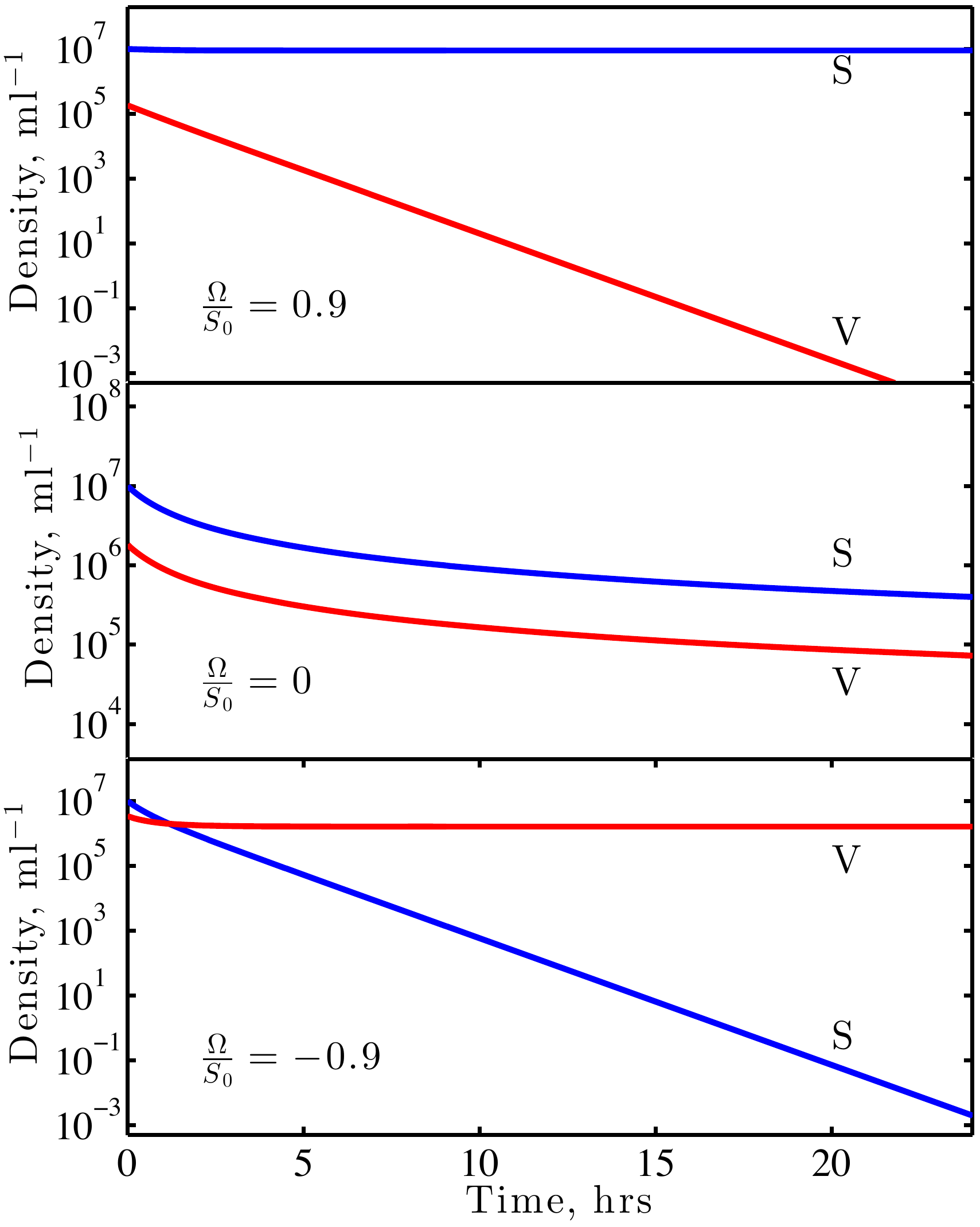}
\caption{Dynamics of susceptible hosts, $S(t)$, and free viruses, $V(t)$,
for the reduced model in Eq.~\req{eq.dsdt}--~\req{eq.dvdt}
in three regimes: $\Omega>0$, $\Omega=0$ and $\Omega<0$ (top, middle
and  bottom, respectively). Common parameters for the dynamics are
$\phi=10^{-7}$ cells/(ml-hrs), $\delta=4.5$ and $S_0=10^7$ cells/ml. 
The initial virus density is $V_0=1.8\times 10^5$,
$1.8\times 10^6$ and $3.5\times 10^6$ viruses/ml in the top, middle
and bottom panels respectively.
\label{fig.regimes}}
\end{center}
\end{figure}

Recall that $(1+\delta)V_0$ is the maximum number of hosts
that can be infected or enter dormancy as a result of interactions
with viruses. Therefore, when $S_0>(1+\delta)V_0$ then there
are enough hosts for all viruses to infect cells ($V_0$ in total)
and to catalyze  $\delta$ hosts per infected cell to enter dormancy ($\delta V_0$
in total).  
This is the case when $\Omega=S_0-(1+\delta)V_0$ is positive.  
In this limit, all viruses
infect a cell, while some hosts remain uninfected.  The condition
$\Omega>0$ represents the
``virus-depletion'' limit.  In contrast, when $S_0<(1+\delta)V_0$
there are not enough hosts for all viruses to infect cells
and to catalyze $\delta$ hosts per infected cell to enter dormancy.
This is the case when $\Omega=S_0-(1+\delta)V_0$ is negative.
In this limit, all hosts are either infected or enter dormancy,
while the viruses remain in the system. The condition
$\Omega<0$ represents the
``host-depletion'' limit.  The condition $\Omega=0$ represents
the critical point dividing these two dynamical regimes
(see Figure~\ref{fig.regimes}).

\begin{table}[t!]
\begin{center}
\begin{tabular}{c|c|c|c}
Variable & $\Omega<0$ & $\Omega=0$ & $\Omega >0$ \\
\hline\hline
$S$  & 0 & 0  & $\Omega$ \\ 
$C$  & 0 & 0  & 0 \\ 
$D$  & $\frac{\delta S_0}{1+\delta}$  & $\delta V_0$   & $\delta V_0$  \\
$I$  & $\frac{S_0}{1+\delta}$  &  $V_0$   &  $V_0$ \\
$V$  & $-\frac{\Omega}{1+\delta}$ & 0   &  0
\end{tabular}
\caption{Asymptotic densities of state variables given
the control parameter $\Omega=S_0-(1+\delta)V_0$.
The conditions $\Omega<0$ and $\Omega>0$ represents
the host-depletion and virus-depletion limits.  See
the text for more details.
\label{tab.asymp}}
\end{center}
\end{table}

\subsection{Viruses can induce nearly all hosts to enter dormancy, even when the virus-host ratio is far less than one}
\begin{figure*}[t!]
\begin{center}
\includegraphics[width=0.45\textwidth]{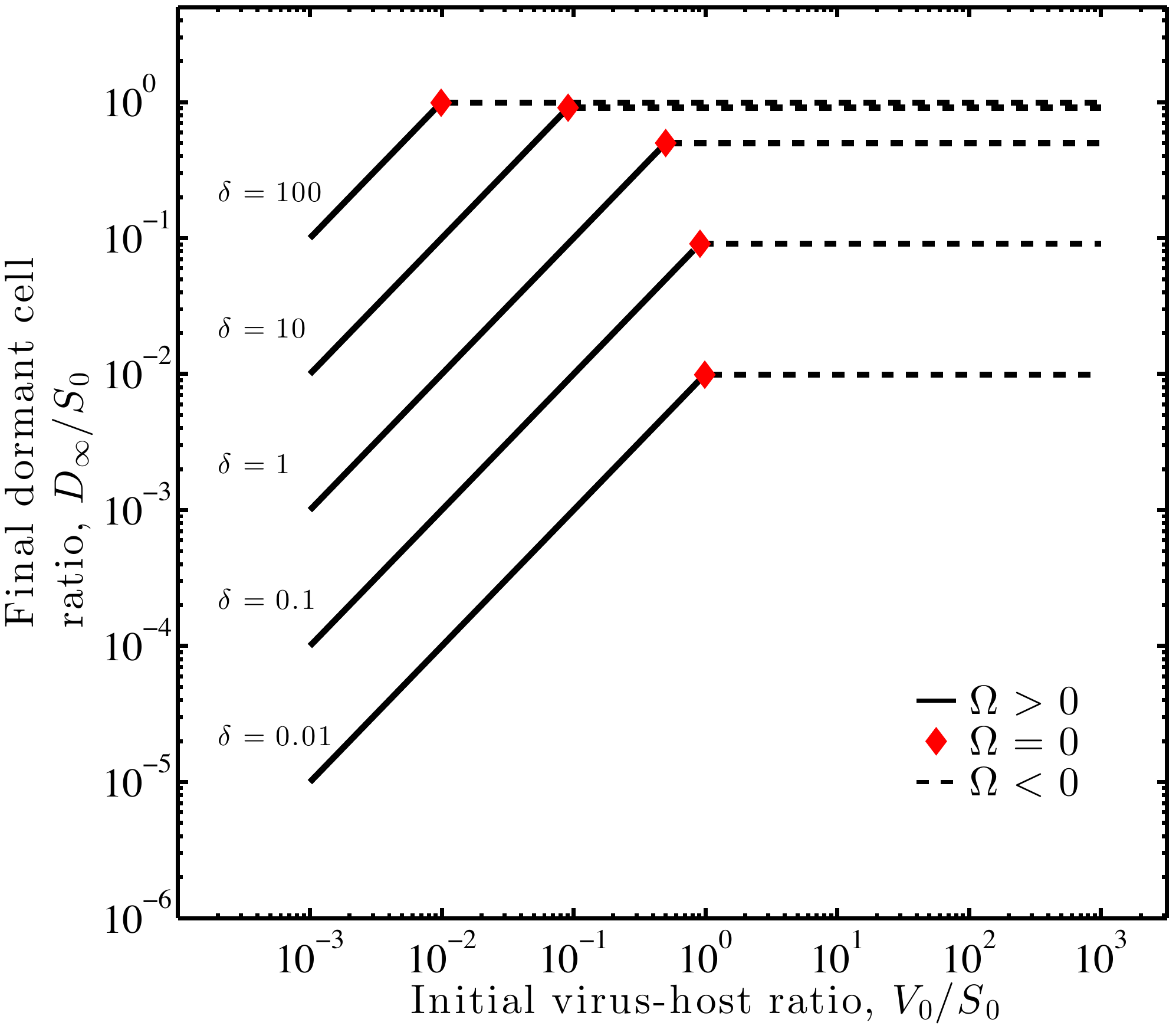}
\mbox{\hspace{0.05\textwidth}}
\includegraphics[width=0.45\textwidth]{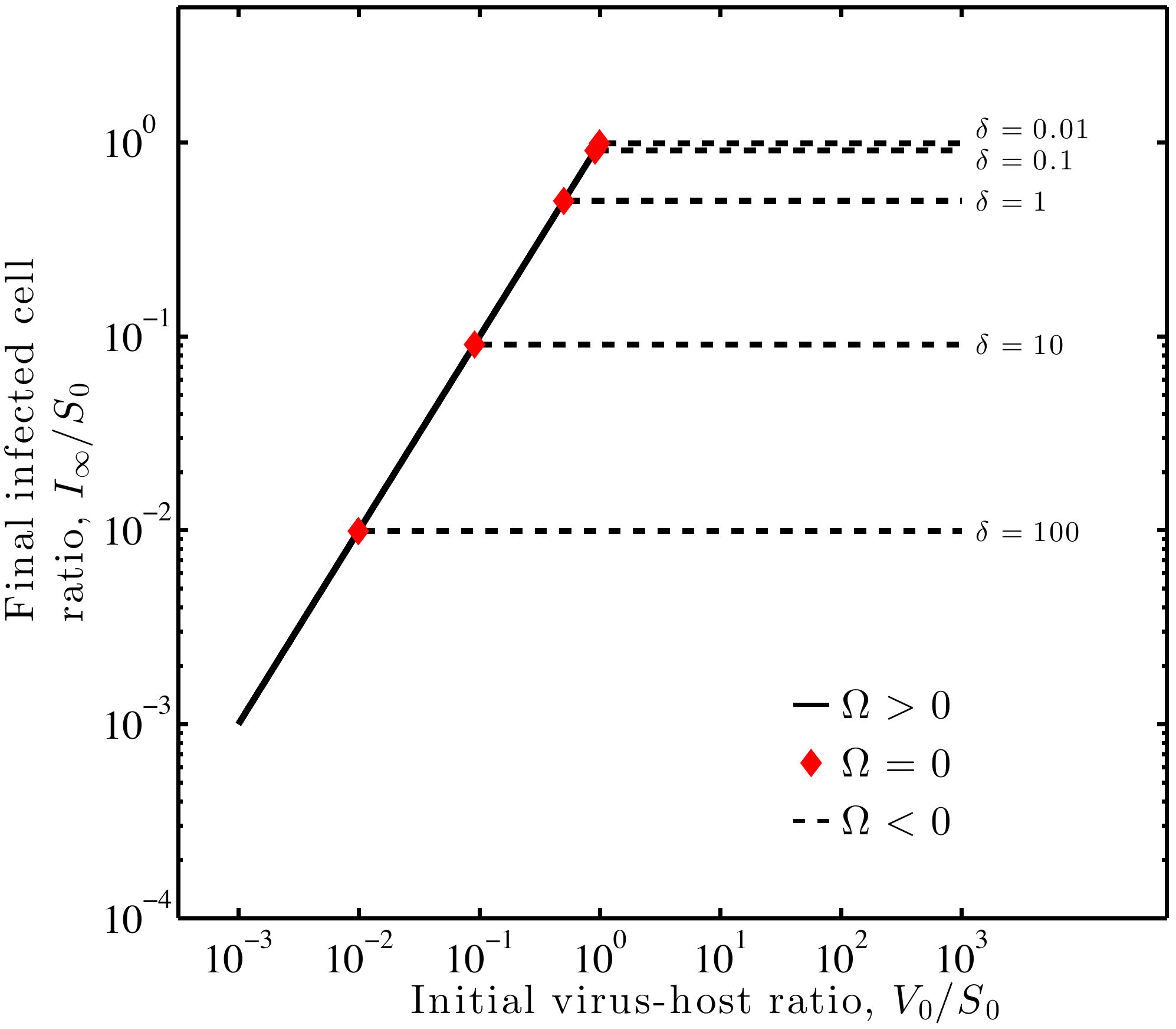}
\caption{Asymptotic fraction of dormant and infected cells resulting from
virus-host dynamics.  (Left) Final dormant cell fraction as a function
of the initial ratio of viruses to hosts, $V_0/S_0$. 
(Right) Final infected cell fraction as a function
of the initial ratio of viruses to hosts, $V_0/S_0$. 
As shown, the final cell state fractions
depend on $\delta$ with qualitatively
different responses as a function of the control parameter
$\Omega$.  
The fraction of dormant cells and infected cells increases
linearly with $V_0/S_0$ so long as
$\Omega>0$.  When $\Omega<0$ then the fraction of dormant and infected
cells is a constant, irrespective of $V_0/S_0$. Increases
in $\delta$ lead to a relative increase in the proportion of
dormant vs.~infected cells.  Parameters are otherwise the
same as in Figure~\ref{fig.regimes}.
\label{fig.dormant}}
\end{center}
\end{figure*}

Traditional analysis of virus-host interactions presupposes that
entrance, through injection or other means, of virus genomes into a host is required for virus-mediated
modification of host cell physiology.  Here, as in traditional models,
$V_0$ represents the upper limit to the
number of hosts infected by viruses (see Table~\ref{tab.asymp}).  
This limit holds when restricting
attention to short-term dynamics before replication and lysis which
releases more viruses
that can initiate subsequent infections.
However, in the present model, hosts can also undergo contact-mediated
dormancy.  When the ratio of viruses to hosts is small, i.e., 
the MOI is ${\cal{M}}_0\equiv V_0/S_0\ll 1$, we find an unexpected outcome:
nearly all of the hosts can enter dormancy even when there are far fewer
viruses than hosts.  In the host-depletion regime, $\Omega<0$, then
$D_{\infty}=\frac{\delta S_0}{1+\delta}$.  If $\delta\gg 1$ then
$D_{\infty}\rightarrow S_0$.  This condition holds 
so long as the relative rates of 
unbinding and dormancy are high relative to infection and
there are enough viruses.  The critical virus density depends on $\delta$
and is equal to $S_0/(1+\delta)$.  For virus densities above this value, then
the dormant cell fraction will have reached its maximum because the
system moves from being host-depleted to virus-depleted.
The comparison of the asymptotic dormant cell fraction and
infected cell fraction are shown in Figure~\ref{fig.dormant}.  As is
apparent, more cells become dormant and infected with increasing
titer.  Yet, the balance of dormancy or infected cell fates shift
with increases in $\delta$.  As $\delta$ increases, then many dormant
cells are initiated for each infected cell, whereas when $\delta$ decreases,
then very few dormant cells are initiated for each infected cell.

\subsection{Dynamics of dormancy induction in the archaeon \emph{Sulfolobus islandicus}}
We apply our biophysical model of contact-mediated dormancy to
a recent empirical study of interactions between
the archaeon \emph{S.~islandicus} and the dsDNA 
fusellovirus \emph{Sulfulobus} spindle shaped virus (SSV9).  
In the experiment, viruses were introduced at relatively
low concentrations compared to that of hosts.
The ratio of viruses to hosts, ${\cal{M}}$ was estimated
to range between 0.01, given plaque-forming unit counts, and 0.01,
given quantitative PCR counts.  
In this experiment nearly 100\% of host cells entered dormancy.
Hence, there was a 10-fold to 100-fold increase
in the conversion of host cells into a dormant state.  SSV9 was
then exposed to UV light to de-activate the virus population.
The subsequent conversion estimates into dormant
cells were statistically unchanged.  We interpret this
result to mean that contact between viruses and host may be
sufficient to induce a population-wide dormancy response
even when viruses are present at densities far below that
of target host cells.  These qualitative results are the basis
for our quantitative parameterization and analysis of the 
model.

The governing parameters of the biophysical model are 
$\phi$, the adsorption rate, and $\delta$, the ratio of
dormancy induction to infection.   Bautista and colleagues
estimated 
$\phi_{original}$ to be 8.4$\times 10^{-11}$ ml/min based on the decay
of plaque-forming units.
They estimated the adsorption rate using
the formula $\phi_{original} = \frac{2.3\log{\frac{V_0}{V_t}}}{N_0 t}$
where $V_0$ is the original titer of viruses, $V_t$ is the
titer at time $t$ and $N_0$ is the original titer of hosts. 
Conventional estimates are that $\phi= \frac{\log{\frac{V_0}{V_t}}}{N_0 t}$,
hence we downward adjust the adsorption rate to be
$\phi = 2.2\times 10^{-9}$ ml/hr (note the change
in units).  The effective adsorption
rate is a combination of the process of diffusion-limited
contact and successful infection. We estimate the
diffusion-limited contact rate based on physical principles standard in the
study of virus-host interactions~\citep{bergpurcell_1977,murray_meps1992}:
\begin{equation}
k_+ \approx 4\times 10^{-9}\frac{r_v}{r_h} \textrm{ml}/\textrm{hr}
\end{equation}
where $r_v$ is the effective radius of the
virus, $r_h$ is the effective radius of the host, where the 
prefactor is appropriate
for interactions taking place at room temperature (293 degrees K)
and in a medium with the viscosity of water.  Assuming
$r_h=1$ $\mu$m and $r_v=0.04$ $\mu$m then we predict $k_+=10^{-7}$
ml/hr.  The ratio of the diffusion-limited contact rate
expected from first principles, $k_+$, and the
realized adsorption rate measured in the experiment, $\phi$ can be
used to estimate $q=\phi/k_+=0.022$.   In this limit, then
$q\approx k_f/k_-$, such that $\delta = p/q$.  
\begin{figure*}[t!]
\begin{center}
\includegraphics[width=0.95\columnwidth]{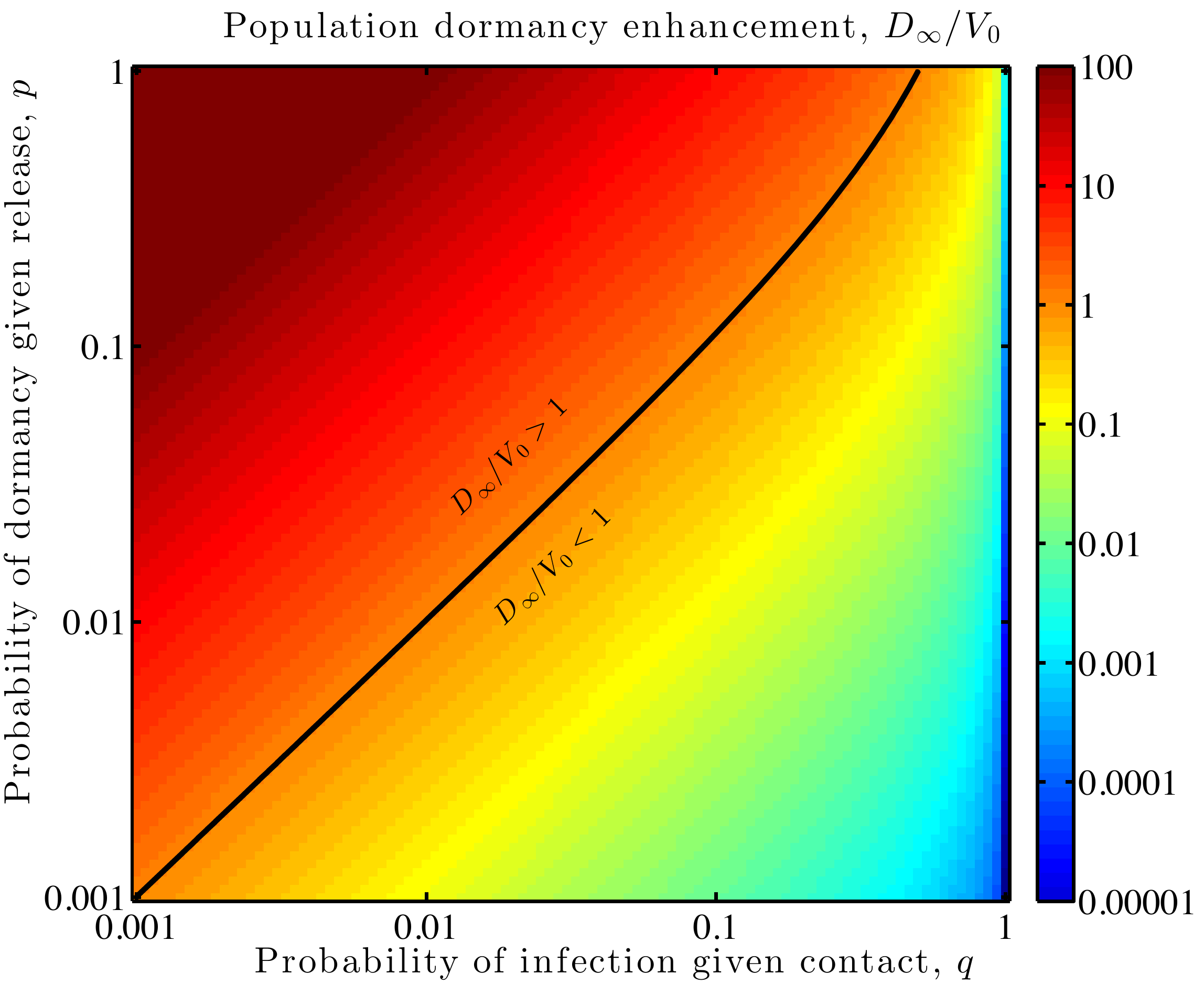}
\mbox{\hspace{0.05\textwidth}}
\includegraphics[width=0.95\columnwidth]{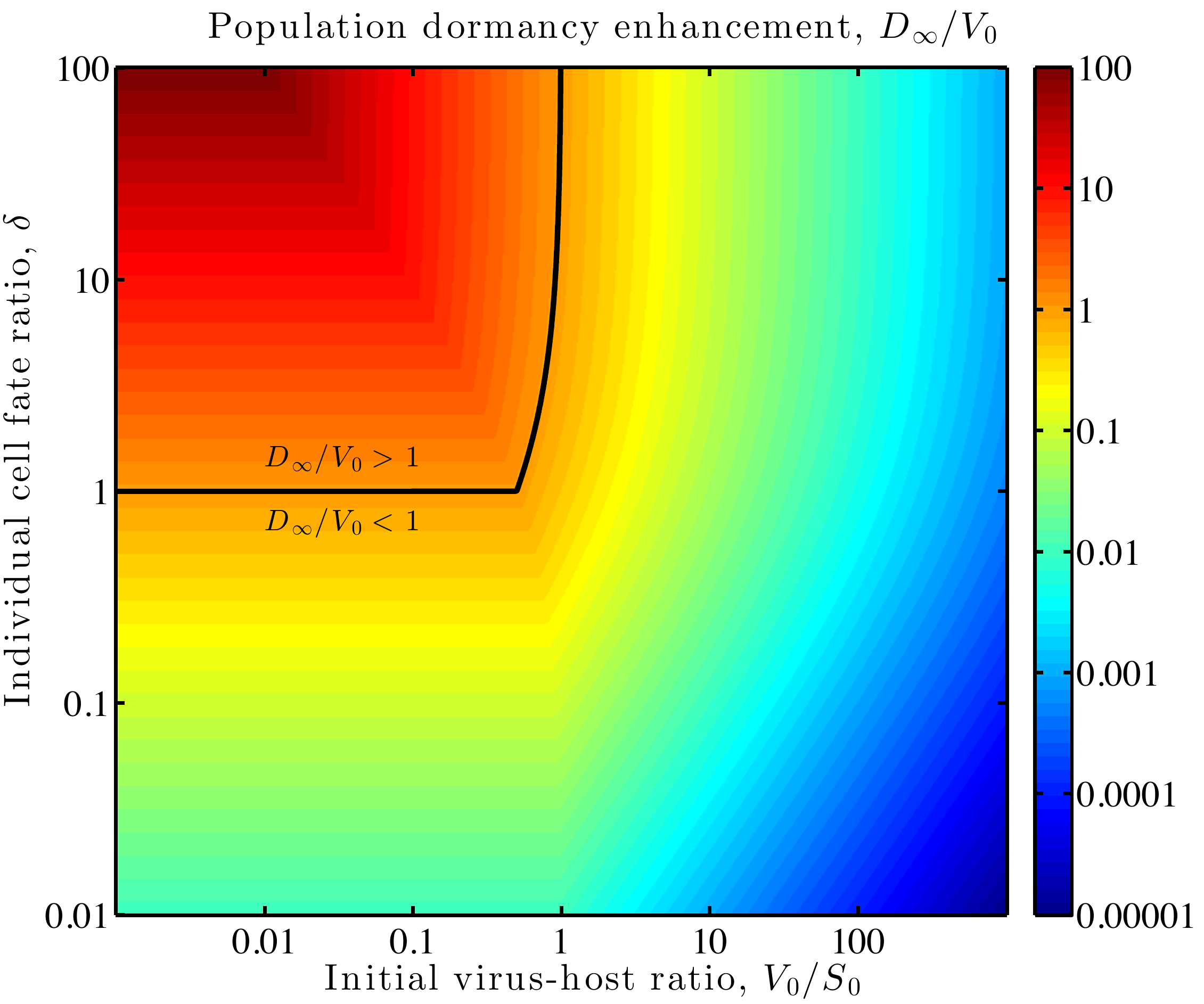}
\caption{Ratio of the concentration of dormant cells at the end of the dynamics to the 
concentration of viruses at the start of the dynamics.
The intensity values denote the ratio, $D_{\infty}/V_0$.  The solid line
demarcates the boundary between ratios that exceed one (upper-left portion) 
and those that are less than one (remainder). 
(Left) Population dormancy enhancement as a function 
of the probability of infection given contact, $q$, and the
probability of dormancy-initiation given reversal of contact, $p$.
Here, $S_0=2.5\times 10^8$ cells/ml and $V_0=S_0/100$.
(Right) Population dormancy enhancement as a function 
of the initial virus-host ratio, $V_0/S_0$,
and the cell fate ratio, $\delta\approx p/q$.
\emph{Key point: the number of cells that enter dormancy
per virus can be much greater than one, even if the
initial virus-host ratio is much smaller than 1.}
\label{fig.enhancement}}
\end{center}
\end{figure*}

We can explore the predicted fraction of
dormant cells as a function of $\delta$.  First, we analyze the ratio
of dormant cells to initial viruses given variation in
$q$ and $p$ (Figure~\ref{fig.enhancement}-left).  For $q\ll 1$
there is a broad range of values of $p$ such that
$D_{\infty}/V_0\gg 1$.  This result means that many hosts
cells could enter dormancy given exposure to low relative titer
of viruses.  Yet, the value of $p$, and therefore of
$\delta$, remains a free parameter given the experimental
tests conducted in the system.  The maximum value of $\delta$
is when $p\rightarrow 1$, i.e., when the conditional probability
of inducing dormancy given a reversible contact approaches 1.
In this case, $\delta\rightarrow 1/0.022=45$.  Lower values of $\delta$
are possible when $p\rightarrow 0$, i.e., when the conditional probability
of inducing dormancy given a reversible contact approaches 0.  In this
limit, $\delta\rightarrow 0$.  The experimental finding of a range
between $\hat{\delta}=10$ to $\hat{\delta}=100$ fold
enhancement is consistent with our finding of
$\delta=45$ and a model in which $p\rightarrow 1$ (see Figure~\ref{fig.enhancement}-right).

We provide another evaluation of the model by considering
the timescale over which contact-mediated
dormancy should take place.  The appropriate time-scale is predicted 
to be
$\tau_c=\frac{1}{\phi(1+\delta)V_0}$.  The approximate cell density
for experiments in~\citep{bautista_inpress} was $S_0\approx 2.5\times 10^8$ ml$^{-1}$. We assume that viruses were present at $V_0\approx 0.01 S_0$.  
Using these values we estimate $\tau_c\approx 4$ hrs.  
Hence, we
predict a characteristic time-scale for conversion of 64\% 
of hosts, corresponding to a one-log drop in susceptible host
density, in a time period of 4 hrs and to conversion of 87\% of
hosts, corresponding to a two-log drop in susceptible
host density, in a time period of 8 hrs.  We view
this time-scale analysis to be another confirmation of the model,
given that even if the hosts initiate dormancy after contact, 
the dormancy ``phenotype'' is likely to be delayed given the
re-organization of intracellular dynamics.  
In summary, non-infectious contacts could happen sufficiently frequently
so as to rapidly induce dormancy on relevant time-scales
of experimental observations.

\section{Discussion}
We have proposed a biophysical model of host-dormancy initiated
by contact with viruses.  The model
explicitly accounts for the possibility that viruses can contact
host cells reversibly.  Reversible contacts may, with some frequency,
lead to induction of dormancy.  Such
contact-mediated dormancy at the cellular scale can be evident
at the population-scale in certain limits. In particular, 
we predict a critical transition to a regime
in which the vast majority of cells become dormant even if the initial
ratio of viruses to hosts is quite small.  This regime is found to
be robust to a broad range of biophysically-relevant parameters.

The inspiration for the model was a recent
series of findings that the majority of an archaeal population could
enter a stasis-like ``dormancy'' in less than 24 hrs after exposure
to a relatively small number of viruses~\cite{bautista_inpress}.  
The same effect
was observed whether active or inactivated viruses were utilized.  
This experimental finding suggests the possibility that contact
between virus particles and host surfaces induce a transformation
in host phenotype.  Dormant cells were unlikely
to be infected and lysed by viruses.  However, such dormancy comes
at a cost, as residing in a dormant state for too long can lead to
loss of cell viability and cell death~\citep{bautista_inpress}.
This transformation
may represent a form of phenotypic plasticity on the part of hosts.
Further work would be required to evaluate whether dormancy
could be initiated independent of virus-contact, which would
represent a form of bet hedging.

We fit our model to the experimental host-virus system, leaving
only one free parameter: the conditional probability of
dormancy initiation upon a reversible contact.  
We predict that whenever this conditional probability is sufficiently
high, then large-scale initiation of dormancy can occur even
when very few cells are infected.  Based on our fits,
we predict rapid initiation of dormancy can take place on a time-scale
of 4-8 hrs, sufficiently fast so as to identify a dormancy phenotype
amongst the majority of the host population in the 12-24 hrs period
as observed.  Moreover, the same model fits predicts the potential
for a 50-fold enhancement in dormant cells with respect to viruses.  
Experiments observe higher enhancement ratios ranging from
10- to 100-fold~\citep{bautista_inpress}.
The uncertainty is due, in part, to challenges in quantifying
virus titer.  Yet, there are other challenges, including
potential errors in the estimation of the adsorption rate and approximation
of the maximum contact rate.  There may also be additional mechanisms
of relevance, e.g., cell-cell communication, as a means
to amplify a small viral contact ``signal'' or
additional spatial structure
of the cells and viruses in the environment not accounted for
in the present model.
Nonetheless, the current
biophysical model provides an explanation for the
qualitative features of dormancy-enhancement
and the time-scale of the effect. We suggests
that the present baseline model serve as the basis for 
future detailed investigations of the phenomena moving forward.

A number of issues remain to link the proposed early-time
dynamics with long-term dynamics.  First, nonlinear feedbacks
are likely to arise in this system due to the infection and release
of viruses.  These viruses are themselves metastable, and so incorporating
the decay of infectious virus particles will also need to be considered.
Second, here we assume that viruses cannot infect dormant cells.  The
interaction between viruses and dormant cells is not fully elucidated.
Finally, it is known that
intracellular interactions of \emph{S.~islandicus} and its viruses are mediated,
in part, by the CRISPR/Cas immune system. 
The CRISPR/Cas immune system is ubiquitous in bacteria
and archaea.  CRISPR/Cas enable host cells to target and degrade 
foreign genetic
elements, including viruses~\citep{barrangou_2007,horvath_2010crispr,vanderoost_2014}.
CRISPR-mediated interactions
can lead, over time, to the diversification of the host as it obtains new immune
elements from the virus and to the diversification of the virus~\citep{andersson_2008virus,isi_:000282210700006,weinberger_2012,held2013reassortment}.
Linking early- to long-term dynamics will also need to confront
the potential diversification of communities arising due
to contact-mediated and infection-mediated dynamics.

In summary, dormancy is a feature of organisms spanning animals to plants to 
microbes.
The evolution of dormancy has long been thought to represent
a way to maximize long-term fitness in an uncertain environment~\citep{cohen_1966}. For microbes,
part of the uncertainty in their fitness
stems from the possibility that they may be infected and lysed
by a virus.  Here, we find that a biophysical mechanism
of context-initiated dormancy
can explain observations of rapid and large-scale conversion
of a host archaeal population into a dormant state by a relatively
small number of viruses.  
We suggest that detailed investigations of contact of hosts
by viruses is likely to yield new biological surprises.
In turn, new theoretical approaches are needed
to consider the integration of ``fast'' dynamics at contact-scales
with the long-term nonlinear feedbacks arising from the effects
of physiological transformations and infection on host and virus populations.

\section*{Acknowledgments}
The authors thank M. Bautista, R. Whitaker and M. Young
for helpful comments and feedback on the manuscript.
JSW acknowledges support from a
Career Award at the Scientific Interface from the
Burroughs Wellcome Fund and NSF Award 1342876.

\appendix

\section{Analytical solution of the contact-mediated dynamics model}
\footnotesize{
\label{app.analytics}
The following reduced model introduced in
Eqs~\req{eq.dsdt}--\req{eq.dvdt}
represents dynamics of susceptible
hosts, $S(t)$, and free viruses, $V(t)$:
\begin{eqnarray}
\frac{\mathrm{d} S}{\mathrm{d} t} &=& -\phi SV(1+\delta) \\
\frac{\mathrm{d} V}{\mathrm{d} t} &=& -\phi SV 
\end{eqnarray}
This model can be solved analytically by observing that:
\begin{equation}
\frac{\mathrm{d} S}{\mathrm{d} V}=(1+\delta)
\end{equation}
such that
\begin{equation}
S(t)=(1+\delta)V(t)+\Omega
\label{eq.sv_relation}
\end{equation}
where the integration constant, $\Omega$, can be solved using
the initial conditions that $S(0)=S_0$ and $V(0)=V_0$:
\begin{equation}
\Omega = S_0 - (1+\delta)V_0
\end{equation}
This two-dimensional system can be reduce to a one-dimensional system
by substituting $S(t)$ as in Eq.~\req{eq.sv_relation}, yielding:
\begin{equation}
\frac{\mathrm{d} V}{\mathrm{d} t}= -\phi V\Bigl((1+\delta)V+\Omega\Bigr)
\label{eq.dvdt_omega}
\end{equation}

When $\Omega\neq 0$, then 
Eq.~\req{eq.dvdt_omega}
can be solved by separation of variables yielding:
\begin{equation}
V(t) = \frac{{\cal{M}}_0 \Omega e^{-\phi\Omega t}}{1-(1+\delta){\cal{M}}_0e^{-\phi \Omega t}}
\end{equation}
where ${\cal{M}}_0$ is the initial population-level multiplicity 
of infection, i.e., the ratio ${\cal{M}}_0=V_0/S_0$.
The solution for $V(t)$ is the basis for a complete
description of the dynamics under the Quasi-steady state approximation:
\begin{eqnarray}
\begin{split}
S(t) &= \frac{\Omega}{1-(1+\delta){\cal{M}}_0e^{-\phi \Omega t}}\\
V(t) &= \frac{{\cal{M}}_0 \Omega e^{-\phi\Omega t}}{1-(1+\delta){\cal{M}}_0e^{-\phi \Omega t}}
\end{split}
\end{eqnarray}
where, in addition:
\begin{eqnarray}
\begin{split}
C(t)&= \frac{\phi S(t) V(t)}{k_f} \\
I(t)&=V_0-C(t)-V(t)\\
D(t)&=\left(S_0-S(t)\right)-\left(V_0-V(t)\right)
\end{split}
\end{eqnarray}
When $\Omega=0$, then 
Eq.~\req{eq.dvdt_omega}
can be solved by integration:
\begin{eqnarray}
\begin{split}
S(t) &= \frac{S_0}{1+S_0\phi t}  \\
V(t) &= \frac{V_0}{1+V_0\phi(1+\delta)t}  
\end{split}
\end{eqnarray}
recalling that when $\Omega=0$ then $(1+\delta)V_0=S_0$.
}

\end{document}